\documentclass[a4paper, aps,pra, twocolumn, superscriptaddress, showpacs]{revtex4}

\usepackage{amssymb}
\usepackage{amsmath}
\usepackage{color}
\usepackage{graphics, graphicx}
\usepackage{bbold}
\usepackage{mathcomp}
\usepackage{subfigure}
\usepackage{verbatim}
\usepackage[colorlinks,citecolor=blue]{hyperref}

\begin{document}

\date{\today}

\title{Topological Fulde-Ferrell states in alkaline-earth-metal-like atoms near an orbital Feshbach resonance}

\author{Su Wang}
\affiliation{Key Laboratory of Quantum Information, University of Science and Technology of China, CAS, Hefei, Anhui, 230026, China}
\affiliation{Synergetic Innovation Center of Quantum Information and Quantum Physics, University of Science and Technology of China, Hefei, Anhui 230026, China}
\author{Jian-Song Pan}
\affiliation{Key Laboratory of Quantum Information, University of Science and Technology of China, CAS, Hefei, Anhui, 230026, China}
\affiliation{Synergetic Innovation Center of Quantum Information and Quantum Physics, University of Science and Technology of China, Hefei, Anhui 230026, China}
\author{Xiaoling Cui}
\affiliation{Beijing National Laboratory for Condensed Matter Physics, Institute of Physics, Chinese Academy of Sciences, Beijing 100190, China}
\author{Wei Zhang}
\email{wzhangl@ruc.edu.cn}
\affiliation{Department of Physics, Renmin University of China, Beijing 100872, China}
\affiliation{Beijing Key Laboratory of Opto-Electronic Functional Materials and Micro-Nano Devices,
Renmin University of China, Beijing 100872, China}
\author{Wei Yi}
\email{wyiz@ustc.edu.cn}
\affiliation{Key Laboratory of Quantum Information, University of Science and Technology of China, CAS, Hefei, Anhui, 230026, China}
\affiliation{Synergetic Innovation Center of Quantum Information and Quantum Physics, University of Science and Technology of China, Hefei, Anhui 230026, China}

\begin{abstract}
We study the effects of synthetic spin-orbit coupling on the pairing physics in quasi-one-dimensional ultracold Fermi gases of alkaline-earth-metal-like atoms near an orbital Feshbach resonance (OFR). The interplay between spin-orbit coupling and pairing interactions near the OFR leads to an interesting topological Fulde-Ferrell state, where the nontrivial topology of the state is solely encoded in the closed channel with a topologically trivial Fulde-Ferrell pairing in the open channel. We confirm the topological property of the system by characterizing the Zak phase and the edge states. The topological Fulde-Ferrell state can be identified by the momentum-space density distribution obtained from time-of-flight images.
\end{abstract}

\maketitle

\section{Introduction}
The recently discovered orbital Feshbach resonance (OFR)~\cite{zhang_orbital_2015,pagano_strongly_2015,hofer_observation_2015} in $^{173}$Yb opens up the possibility of studying strongly interacting phenomena in quantum gases of alkaline-earth-metal and alkaline-earth-metal-like atoms, where the separation of the electronic and nuclear degrees of freedom in the long-lived clock states has led to many potential applications in quantum metrology, quantum information, and quantum simulation~\cite{wu_exact_2003,takamoto_optical_2005,ludlow_systematic_2006,fukuhara_degenerate_2007,gorshkov_alkaline-earth-metal_2009,zhang_spectroscopic_2014,cappellini_direct_2014,scazza_observation_2014,bloom_optical_2014,ludlow_optical_2015}. In an OFR, the atoms in the electronic ground state $^1S_0$ (labeled by $|g\rangle$) interact with those in the long-lived excited $^3P_0$ state ($|e\rangle$) via magnetically tunable interactions. Due to the decoupling of electronic and nuclear degrees of freedom, the two-body interaction at short range occurs either in the electronic spin-singlet channel, with the nuclear spins in the spin-triplet channel, or in the electronic spin-triplet channel, with the nuclear spins in the spin-singlet channel. These short-range interaction channels are further coupled in the presence of a finite magnetic field, which gives the resonant scattering with nuclear-spin exchange processes~\cite{zhang_orbital_2015,pagano_strongly_2015,hofer_observation_2015}.

The combination of versatile quantum control techniques and unique resonant interactions in cold alkaline-earth-metal-like atoms can lead to nontrivial many-body physics~\cite{zhang_kondo_2016,xu_reaching_2016,chen_polarons_2016}. One interesting possibility is to study the effects of synthetic spin-orbit coupling (SOC) close to an OFR~\cite{wall_synthetic_2016,livi_synthetic_2016,song_spin-orbit_2016,kolkowitz_spinorbit-coupled_2017,zhou_symmetry-protected_2016, iemini_majorana_2017}. Raman-induced synthetic SOCs in alkaline atoms have been extensively studied in recent years~\cite{galitski_spin-orbit_2013,zhang_fermi_2014,goldman_light-induced_2014,zhai_degenerate_2015,yi_pairing_2015}. In particular, theoretical studies reveal that SOC in various forms tend to enhance pairing in Fermi gases in two and three dimensions, and may stabilize a topological nontrivial superfluid phase with finite center-of-mass momentum, known as a topological Fulde-Ferrell (tFF) state~\cite{chen_inhomogeneous_2013, qu_topological_2013,zhang_topological_2013, liu_topological_2013}. Meanwhile, the experimental realization of such an exotic pairing state is hindered by the experimental difficulty of heating introduced by the Raman lasers, which typically need to be tuned near resonance with the excited states. The heating issue can be largely overcome in lanthanide atoms such as dysprosium, where the narrow optical transition reduces heating in the Raman process~\cite{cui_synthetic_2013,burdick_long-lived_2016}. Alternatively, in alkaline-earth-metal or alkaline-earth-metal-like atoms, laser-induced heating can also be drastically reduced, either by directly coupling the clock states which serve as pseudospins, or by Raman coupling two different nuclear spin states in the $^1S_0$ manifold via the $^3P_1$ state, whose line width is relatively narrow. Indeed, with the recent experimental realization of SOCs in alkaline-earth-metal or alkaline-earth-metal-like atoms~\cite{wall_synthetic_2016,livi_synthetic_2016,song_spin-orbit_2016,kolkowitz_spinorbit-coupled_2017}, these systems have become the ideal platforms to study SOC physics at low temperatures~\cite{zhou_symmetry-protected_2016, iemini_majorana_2017}.

\begin{figure}
\begin{centering}
\includegraphics[width=1\columnwidth]{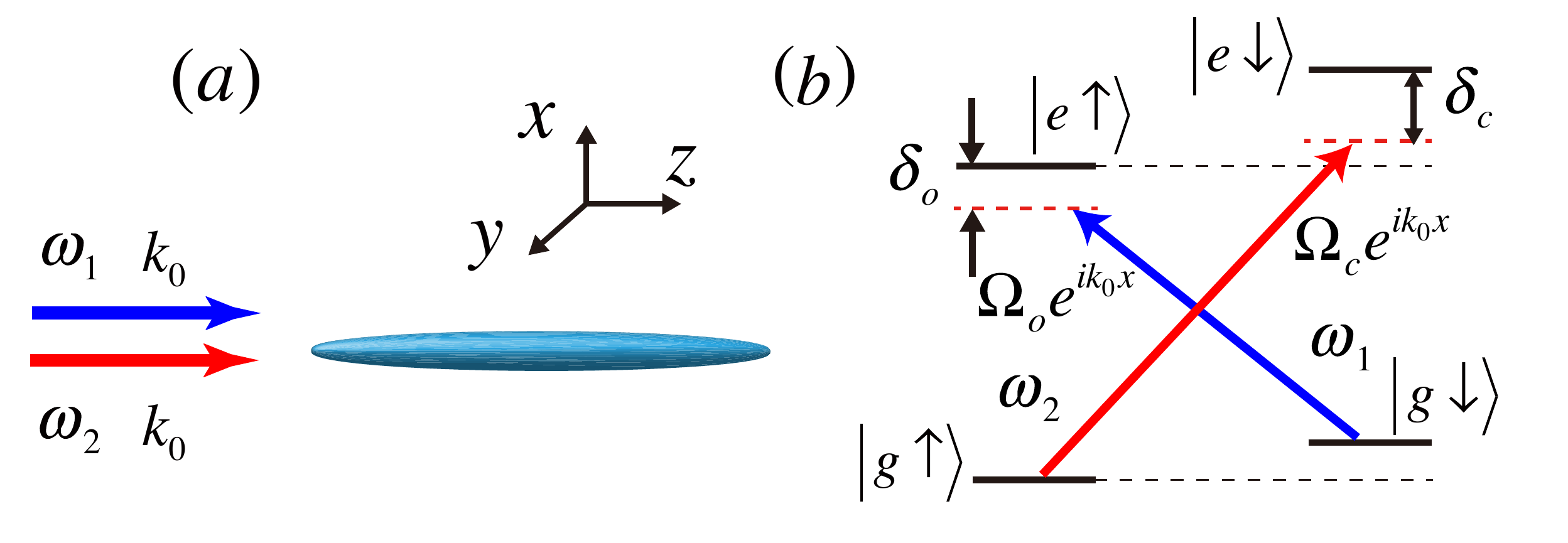}
\par\end{centering}
\caption{\label{fig:Scheme}Schematic illustration of the system setup. (a) A quasi-one-dimensional atomic gas is
driven by two lasers that are right- and left-circularly polarized, respectively. The right- (left-) circularly polarized laser has the frequency
$\omega_{1}$ ($\omega_{2}$) and the wave vector $k_{0}$. The quantization
direction is set as the $z$ direction. (b) A simplified level scheme showing the relevant clock states. Other hyperfine states in the clock-state manifold can be made far-detuned via a differential Stark shift~\cite{mancini_observation_2015,song_spin-orbit_2016}.}\label{fig:fig1}
\end{figure}

In this work, we study the pairing physics in spin-orbit coupled alkaline-earth-metal-like atoms near an OFR. We focus on a quasi-one-dimensional setup, where it has been shown previously that the interplay of SOC, $s$-wave interaction, and effective Zeeman fields can lead to topologically nontrivial pairing states in alkaline atoms. With these elements in mind, we consider a cross-coupling scheme as illustrated in Fig.~\ref{fig:fig1}, where right- (left-) circularly polarized light couples the $|g\downarrow\rangle$ and $|e\uparrow\rangle$ ($|g\uparrow\rangle$ and $|e\downarrow\rangle$) states. Here, $|g\uparrow\rangle$ ($|e\uparrow\rangle$) and $|g\downarrow\rangle$ ($|e\downarrow\rangle$) correspond to two different nuclear-spin states in the $^1S_0$ ($^3P_0$) manifold.

In the absence of SOC, the interorbital spin-exchange nature of the OFR interaction typically leads to two separate pairing order parameters in the open ($\{|g\downarrow\rangle,|e\uparrow\rangle\}$) and the closed ($\{|g\uparrow\rangle,|e\downarrow\rangle\}$) channels, respectively. Adopting a mean-field approach~\cite{pethick_bose-einstein_2002}, we find that under the SOCs considered here, the Bogoliubov quasiparticle spectra for the open and the closed channels are decoupled. However, contributions from the two channels are coupled in the thermodynamic potential, which in turn affect the many-body ground state. This unique feature of the OFR gives rise to an interesting topological Fulde-Ferrell state, where the external magnetic field induces a finite center-of-mass momentum for both the closed- and open-channel order parameters. Within the experimentally relevant parameter regime that we have considered, we find that the nontrivial topology of the tFF state is solely encoded in the closed channel. The topological property of the system is characterized by the Zak phase as well as the edge states in the energy spectrum, in which the open- and closed-channel contributions can be separately identified. We further investigate the stability of the tFF state by mapping out the phase diagram and propose to detect the tFF state from its signatures in the momentum-space density distributions. Our results can serve as a first step in the understanding of interesting pairing physics in spin-orbit coupled, strongly interacting alkaline-earth-metal or alkaline-earth-metal-like atoms.

The remainder of the paper is organized as follows. In Sec.~\ref{sec:model}, we introduce the model Hamiltonian and the mean-field approach corresponding to the setup in Fig.~\ref{fig:fig1}. In Sec.~\ref{sec:tff}, we study in detail the tFF state. We then map out the typical phase diagram of the system in Sec.~\ref{sec:diagram}, and discuss a possible detection scheme of the tFF state based on the number density distribution in Sec.~\ref{sec:tof}. Finally, we summarize in Sec.~\ref{sec:summary}.

\section{Model}
\label{sec:model}
The many-body Hamiltonian corresponding to the configuration in Fig.~\ref{fig:fig1} can be written as
\begin{equation}
  H=\sum_{k, i}\phi^\dag_{i,k} H^{(i)}_{0,k} \phi_{i,k}+H_{\rm int},
\end{equation}
where $i=(o,c)$ labels the open- and the closed-channel contributions, and
\begin{eqnarray}
H^{(i)}_{0,k}&=&\left(\begin{array}{cc}
\frac{\hbar^2}{2m}\left(k-\frac{k_0}{2}\right)^{2}+\delta_{i} & -\hbar\Omega_{i}\\
-\hbar\Omega_{i}^{*} & \frac{\hbar^2}{2m}\left(k+\frac{k_0}{2}\right)^{2}
\end{array}\right),\\ \nonumber
H_{\rm int}&=&\frac{1}{2L}\sum_q\left(g_{+}{A}_{+,q}^{\dagger}{A}_{+,q}+g_{-}{A}_{-,q}^{\dagger}{A}_{-,q}\right).
\end{eqnarray}
Here, $\phi_{o,k}^{\dagger}=\left(\begin{array}{cc} \psi_{e\uparrow, k}^{\dagger} & \psi_{g\downarrow, k}^{\dagger}\end{array}\right)$ and $\phi_{c,k}^{\dagger}=\left(\begin{array}{cc} \psi_{e\downarrow, k}^{\dagger} & \psi_{g\uparrow, k}^{\dagger}\end{array}\right)$ with $\psi_{\lambda, k}$ ($\psi_{\lambda, k}^{\dagger}$) the annihilation (creation) operator for the state labeled by $\lambda=\{g\uparrow,g\downarrow,e\uparrow,e\downarrow\}$, $\delta_o$ ($\delta_c$) and $\Omega_o$ ($\Omega_c$) are respectively the laser detuning and Rabi frequency in the open (closed) channel, and $L$ is the quantization length. For the interaction Hamiltonian, $A_{\pm,q}=\sum_k\big(\psi_{e\downarrow,q-k}\psi_{g\uparrow,k}\mp\psi_{e\uparrow,q-k}\psi_{g\downarrow,k}\big)$, and the effective one-dimensional interactions $g_{+}$ and $g_{-}$ are related to the scattering lengths in three dimensions and can be tuned either by external magnetic field via the OFR or by transverse trapping frequency via the confinement-induced resonance~\cite{zhang_kondo_2016}. Note that in writing down the Hamiltonian above, we have assumed that the motional degrees of freedom in the tightly confined transverse directions are frozen and have been integrated out. Importantly, under the coupling lasers, in order for the spin-exchange interaction to be time independent, we must have $\delta_c-\delta_o=\delta$, where the differential Zeeman shift $\delta=\omega_{e\downarrow}-\omega_{e\uparrow}-\left(\omega_{g\downarrow}-\omega_{g\uparrow}\right)$.

By defining the pairing order parameters $\Delta_{\pm}=g_{\pm}\langle A_{\pm,Q}\rangle/2L$, as well as the open- and closed-channel order parameters $\Delta_o=\Delta_-+\Delta_+$ and $\Delta_c=\Delta_--\Delta_+$, respectively, we can write down the mean-field Hamiltonian in the grand canonical ensemble as
\begin{align}
  H-\mu N = &  \frac{1}{2}\sum_{i,k}\tilde{\phi}_{i,k}^{\dagger}\tilde{H}^{\left(i\right)}_{k}\tilde{\phi}_{i,k}+E_0, \label{eqn:Hfull_op}
\end{align}
where $\tilde{\phi}_{o,k}=\left(\begin{array}{cccc}
  \psi_{e\uparrow, k} & \psi_{g\downarrow, k} & \psi_{e\uparrow,Q-k}^{\dagger} & \psi_{g\downarrow,Q-k}^{\dagger}\end{array}\right)^{T}$,
$\tilde{\phi}_{c,k}=\left(\begin{array}{cccc}
\psi_{e\downarrow, k} & \psi_{g\uparrow, k} & \psi_{e\downarrow,Q-k}^{\dagger} & \psi_{g\uparrow,Q-k}^{\dagger} \end{array}\right)^{T}$
, $N=\sum_{i,k}\phi_{i,k}^{\dagger}\phi_{i,k}$, and
\begin{widetext}
\begin{align}
  \tilde{H}^{\left(i\right)}_{k}&=\left(\begin{array}{cccc}
\frac{\hbar^{2}}{2m}\left(k-\frac{k_{0}}{2}\right)^{2}+\delta_{i}-\mu & -\hbar\Omega_{i} &  & -\Delta_{i}\\
-\hbar\Omega_{i}^{*} & \frac{\hbar^{2}}{2m}\left(k+\frac{k_{0}}{2}\right)^{2}-\mu & \Delta_{i}\\
 & \Delta_{i}^{*} & -\frac{\hbar^{2}}{2m}\left(Q-k-\frac{k_{0}}{2}\right)^{2}-\delta_{i}+\mu & \hbar\Omega_{i}^{*}\\
-\Delta_{i}^{*} &  & \hbar\Omega_{i} & -\frac{\hbar^{2}}{2m}\left(Q-k+\frac{k_{0}}{2}\right)^{2}+\mu
\end{array}\right),\label{eqn:Hfull}\\
  E_0&=\sum_{i,k} \left\{ \frac{\hbar^{2}}{2m}\left[\left(k-Q\right)^{2}+\left(\frac{k_{0}}{2}\right)^{2}\right]+\frac{\delta_{i}}{2}-\mu\right\}  -\frac{L}{2g_{+}}\left|\Delta_{c}-\Delta_{o}\right|^{2}-\frac{L}{2g_{-}}\left|\Delta_{c}+\Delta_{o}\right|^{2} . \label{eqn:thermoBare}
\end{align}
\end{widetext}
Note that in a strictly one-dimensional system, the mean-field formalism above is invalid, due to the lack of long-range
order. However, as we are considering a quasi-one-dimensional gas, the residue degrees
of freedom in the tightly confined transverse direction effectively suppress quantum fluctuations. The mean-field description should give a qualitatively valid picture at zero temperature and has been widely applied in this context~\cite{bakhtiari_spectral_2008,chen_inhomogeneous_2013}. Note also that in Eq.~(\ref{eqn:Hfull_op}), we have taken a single center-of-mass momentum $Q$. This is justified by Bogoliubov--de Gennes--type calculations~\cite{de_gennes_superconductivity_1999}, from which we find that the ground state only has a common center-of-mass momentum in both $\pm$ channels. It follows that the center-of-mass momenta of the pairing order parameters in the open and the closed channels are the same. We note that the lack of Larkin-Ovchinnikov--type pairing states, i.e., states whose pairing order parameter features a superposition of $Q$ and $-Q$ center-of-mass momenta, originates from the finite differential Zeeman shift $\delta$, which explicitly breaks the inversion symmetry in the single-particle dispersions and makes the $Q$ and $-Q$ pairing energetically inequivalent.

It is then straightforward to diagonalize the mean-field Hamiltonian and evaluate the thermodynamic potential at zero temperature,
\begin{equation}
  \Omega  =  -\frac{1}{2}\sum_{i,n,k}E_{i,n,k} \Theta\left(E_{i,n,k}\right)+E_0,\label{eqn:thermo}
\end{equation}
where $E_0$ is given by Eq. (\ref{eqn:thermoBare}), $\Theta(x)$ is the Heaviside step function, and the quasiparticle ($n=1,2$) and quasihole ($n=3,4$) dispersions $E_{i,n,k}$ of the open and the closed channels can be calculated by diagonalizing the Hamiltonian in Eq.~(\ref{eqn:Hfull}). The ground state of the system can be determined by minimizing the thermodynamic potential above. For our numerical calculations, we set
$k_0=2\pi\left(556{\,\rm{nm}}\right)^{-1}$, and $\omega_{\perp}=100$ kHz. We use $\omega_{\perp}$ and $k_{\perp}=\sqrt{2m\omega_{\perp}/\hbar}$ as the unit of energy and momentum, respectively.

We characterize the topological properties of the system by calculating the Zak phase~\cite{zak_berrys_1989,atala_direct_2013} of the ground state according to the definition
\begin{equation}
\gamma=-i\int dk\left\langle u_{k}\right|\frac{\partial}{\partial k}\left|u_{k}\right\rangle ,
\end{equation}
where $\left|u_{k}\right\rangle $ is the eigenvector of the occupied bands.
The Zak phase can be either $0$ or $\pi$ according to the definition above,
characterizing a topologically trivial or a nontrivial phase, respectively.

From the formalism above, an important observation is that in the mean-field Hamiltonian Eq.~(\ref{eqn:Hfull}), the open- and the closed-channel contributions are decoupled. This would lead directly to decoupled open-channel and closed-channel quasiparticle spectra and eigen states, which make it possible to identify and differentiate open-channel and closed-channel contributions in the Zak phase. The open-channel and the close-channel contributions to the thermodynamic potential in Eq.~(\ref{eqn:thermo}), however, are coupled in the last two terms. Therefore, the pairing order parameters and the many-body ground state still needs to be solved self-consistently by considering both channels.

\section{Topological Fulde-Ferrell state}
\label{sec:tff}
\begin{figure}
\includegraphics[width=1\columnwidth]{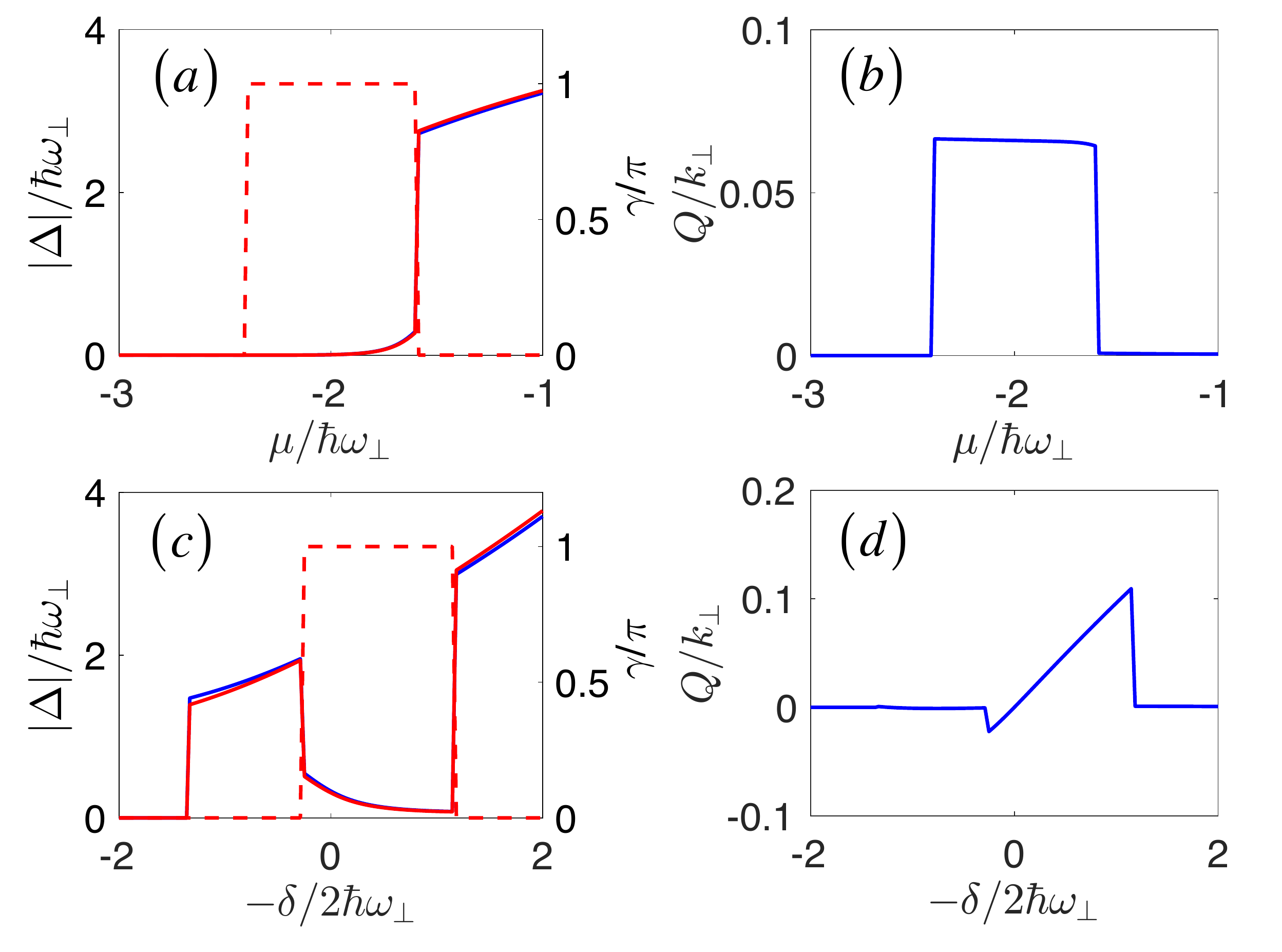}
\caption{(a) Variation of the ground-state order parameters $|\Delta_o|$ (red solid), $|\Delta_c|$ (blue solid), and the Zak phase $\gamma/\pi$ (red dashed) as functions of $\mu$. The left $y$ axis labels the pairing order parameters, and the right $y$ axis labels Zak phase. (b) The ground-state FF momentum as a function of $\mu$. (c) Variation of the ground-state order parameters and the Zak phase as functions of $\delta$. The conventions here are the same as those in (a). (d) The ground-state FF momentum as a function of $\delta$. For (a,b), $\delta\approx -1.35\hbar\omega_{\perp}$, and for (c,d), $\mu\approx -1.70\hbar\omega_{\perp}$. Here, $\Omega_{o}=1.2\omega_{\perp}$, $\Omega_{c}=2.5\omega_{\perp}$, where the units of energy $\omega_{\perp}$ and momentum $k_{\perp}$ are defined in the main text.}\label{fig:Order-parameters}
\end{figure}

We study the ground state of the system as the parameters $\delta$ and $\mu$ are tuned. While changing $\delta$ corresponds to tuning the magnetic field, decreasing $\mu$ corresponds to decreasing the total number density. In Fig.~\ref{fig:Order-parameters}, we show the variation of the order parameters $\Delta_o$, $\Delta_c$, as well as the center-of-mass momentum $Q$. The various abrupt jumps in $\Delta$ and $Q$ indicate the existence of first-order phase transitions, which are the result of the competition between pairing and the effective Zeeman fields induced by the detunings $\delta$. By calculating the Zak phase of the system, we find that topologically nontrivial states exist in the intermediate parameter regimes for both $\delta$ and $\mu$. The topological pairing state features a finite $Q$, which indicates that it is the tFF state previously discussed in alkaline atoms under synthetic SOCs. Interestingly, for the tFF state here, the Zak phase comes from the closed channel only. Thus, at least on the mean-field level, the tFF state here can be regarded as a coherent mixture of tFF pairing in the closed channel and trivial FF pairing in the open channel, with the overall topology of the system determined entirely by the closed channel. We note that in Fig.~\ref{fig:Order-parameters} and for the rest of the paper, we choose the parameters $\Omega_{o}=1.2\omega_{\perp}$ and $\Omega_{c}=2.5\omega_{\perp}$. The stability region of the tFF state will become smaller and eventually vanish if the ratio $\Omega_c/\Omega_o$ decreases toward $1$. Further, as the matrix elements for the clock-state transitions are quite small, given the typical value of $\omega_{\perp}$, our choice of parameters should fall within the experimentally relevant regime for the preparation of the tFF state. We have checked that for different choices of the parameters $\Omega_o$ and $\Omega_c$ within such a regime, our results regarding the tFF state do not change qualitatively.

\begin{figure}
\includegraphics[width=1\columnwidth]{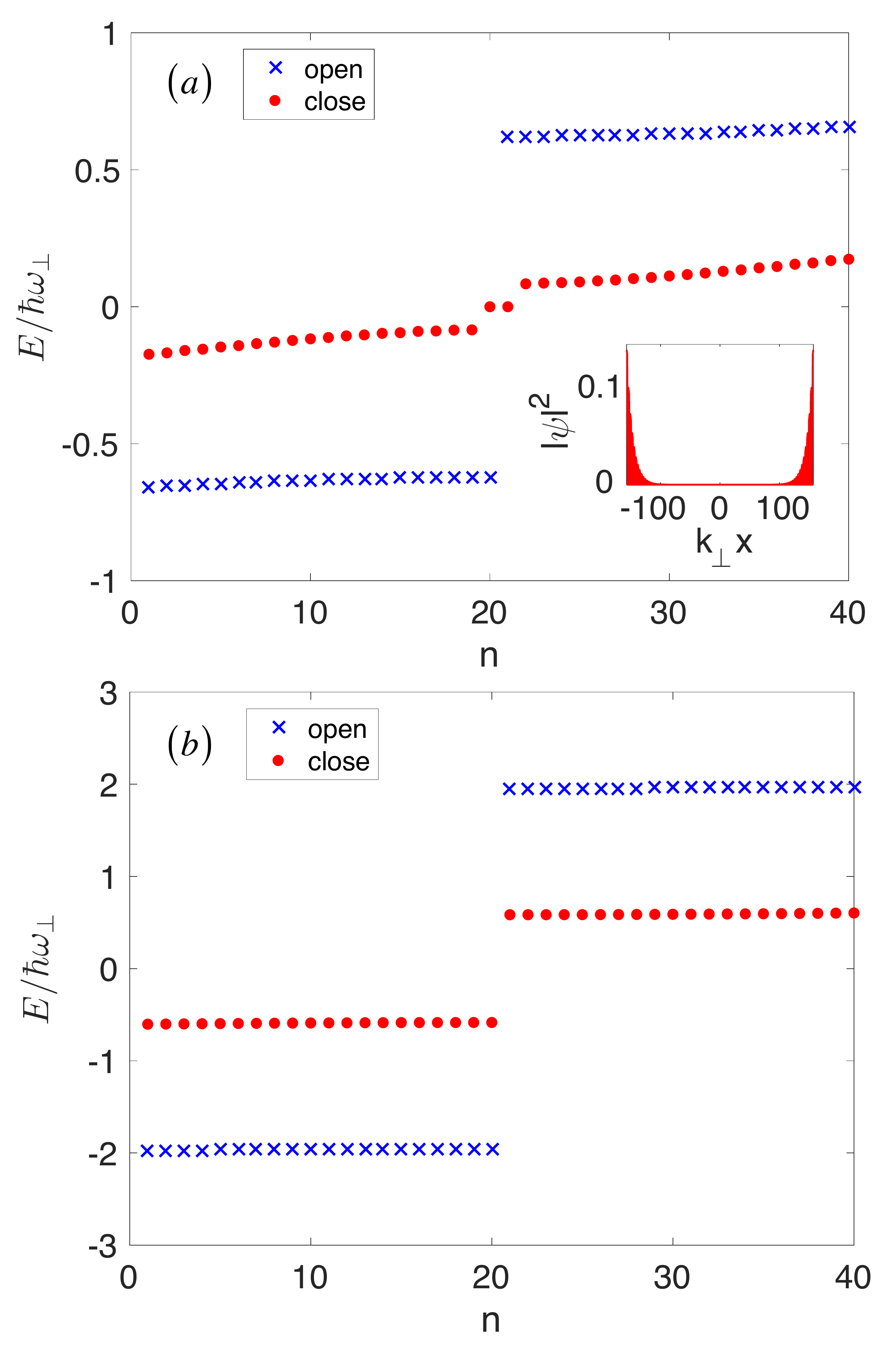}
\caption{\label{fig:Edge-state} Typical energy spectra of (a) the tFF state and (b) the FF state under an open boundary condition. Here, the spectrum of the open (blue crosses) and the closed channels (red dots) are decoupled on the mean-field level. For the tFF state in (a), $\mu=-1.6\hbar\omega_{\perp}$, $\delta=-0.4\hbar\omega_{\perp}$; for the FF state in (b), $\mu=-\hbar\omega_{\perp}$, $\delta=-0.4\hbar\omega_{\perp}$. }
\end{figure}

The topological nature of the tFF state can be further confirmed by characterizing the edge states in a system with open boundary conditions. In Fig.~\ref{fig:Edge-state}, we show the energy spectrum for both the tFF and the trivial FF state for such an open-boundary system. While there are
no zero-energy edge states for the trivial FF phase, a pair of topological edge states emerge in the bulk gap for the tFF phase. Furthermore, as the eigenstates of the open and closed channels are decoupled, in Fig.~\ref{fig:Edge-state}(a) we see the edge states emerge only in the closed-channel section of the tFF phase, consistent with the Zak-phase calculations. The spatial density distributions of the edge states are shown in the inset of Fig.~\ref{fig:Edge-state}(a), which are indeed localized at the two edges.

\section{Phase diagram}
\label{sec:diagram}
\begin{figure}[tbp]
\includegraphics[width=1\columnwidth]{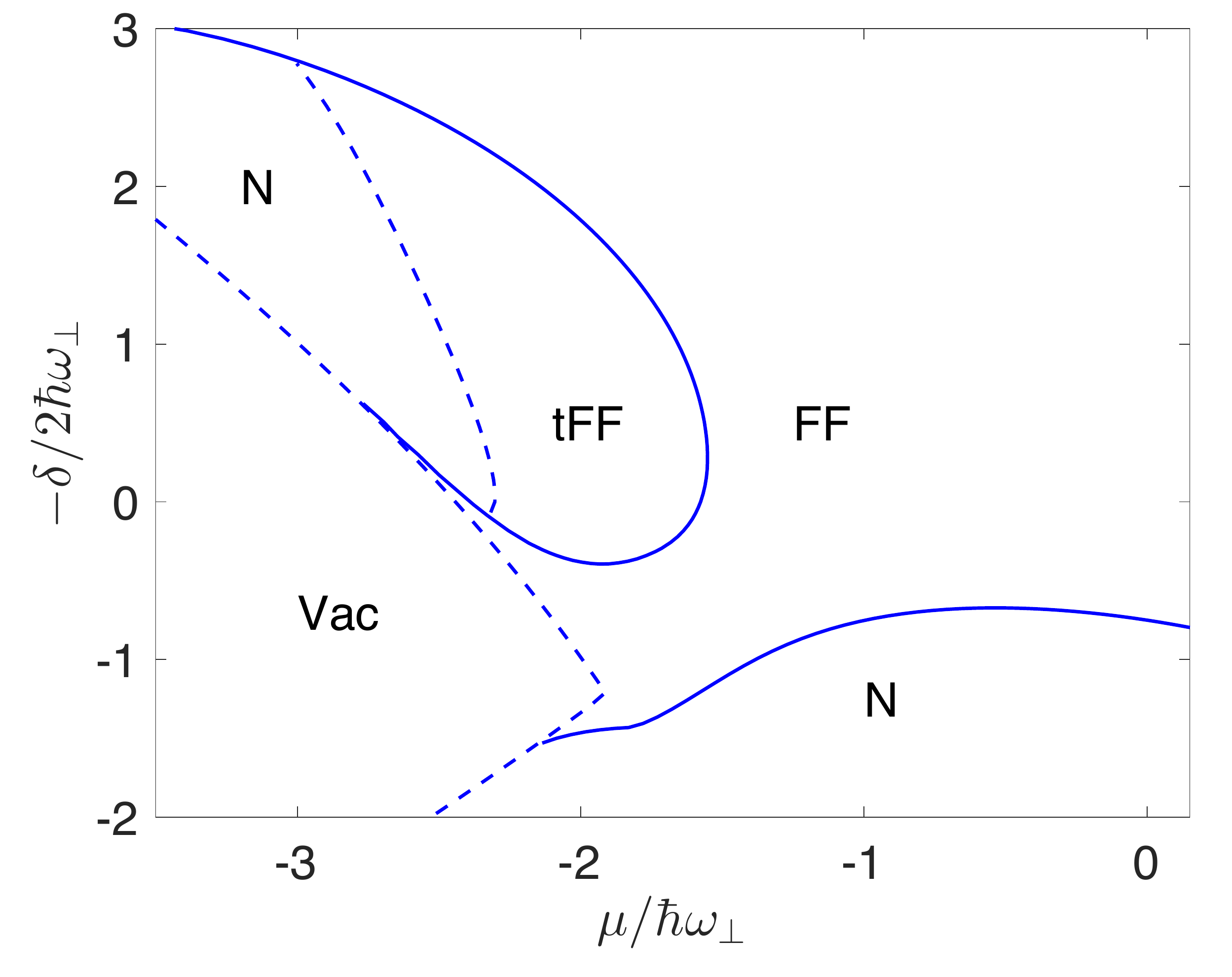}
\caption{\label{fig:Phase-diagrams}Ground-state phase diagram. There are four phases in
the phase diagram: the normal phase (N), the vacuum state (Vac),
the Fulde-Ferrell phase (FF), and the topological Fulde-Ferrell phase
(tFF). The ground state of the system undergoes first-order (second-order) phase transitions across the solid (dashed) boundaries. Here, $\Omega_{o}=1.2\omega_{\perp}$, $\Omega_{c}=2.5\omega_{\perp}$.}
\end{figure}

We show a typical phase diagram of the system on the $\mu$-$\delta$ plane in Fig.~\ref{fig:Phase-diagrams}. The tFF state is stable over a considerable parameter regime against the normal (N) and the topologically trivial FF state. While the tFF-FF phase boundary is of the first order, the tFF-N phase boundary is continuous.

As illustrated in Fig.~\ref{fig:Order-parameters}, a notable difference between the tFF and the FF states is that while the center-of-mass momentum $Q$ of the pairing fields in the tFF state is typically one-tenth of the Fermi wave vector, it is typically vanishingly small for the FF state. Though the feature itself is not directly related to the topological nature of the tFF state, the existence of a nontrivial topology requires that the order parameter $\Delta$ should not be too large, as otherwise the system would be in the BEC regime and becomes topologically trivial. On the other hand, for the SOC-induced FF state here, the finite $Q$ originates from the Fermi-surface deformation induced by the interplay of SOC and effective Zeeman fields. With other parameters fixed, when the magnitude of the pairing field $\Delta$ increases, the Fermi-surface deformation is less important, which tends to result in smaller $Q$. Thus, in the tFF state, $|\Delta|$ is small and $Q$ is large, while in the FF state, the situation is the opposite.

%{\color{red}Add a qualitative discussion about phase diagrams with varying $\Omega$?}

\section{Number density and detection}

\label{sec:tof}
\begin{figure}[tbp]
\includegraphics[width=1\columnwidth]{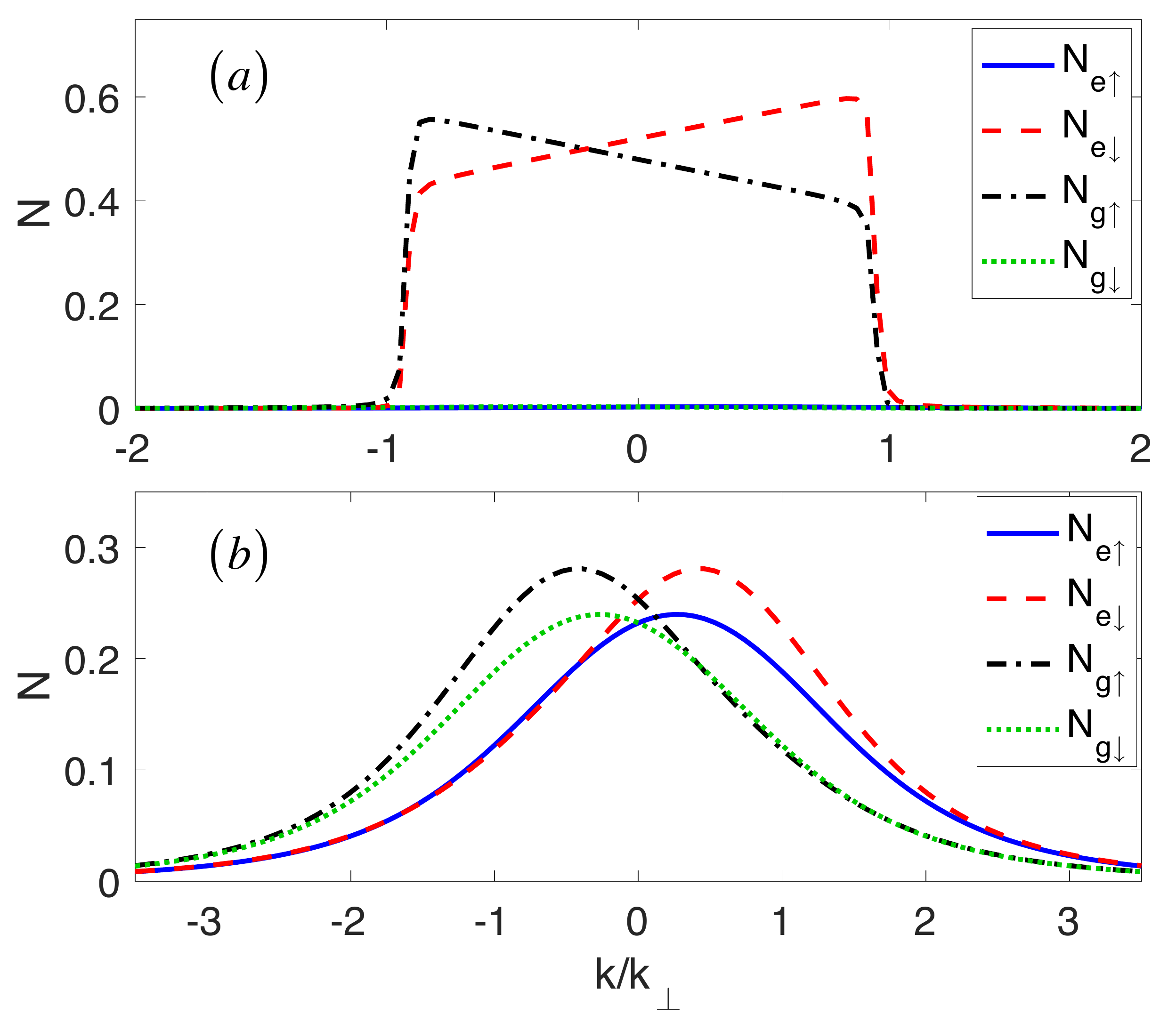}
\caption{\label{fig:Particle-number}The momentum-space density
distribution of a typical (a) tFF state and (b) FF state. The parameters are the same as Fig.~\ref{fig:Edge-state}.}
\end{figure}

The difference in the pairing-field magnitudes of the tFF and the FF states can give rise to discernible features in the momentum-space number distributions of the two states. As shown in Figs.~\ref{fig:Particle-number}(a) and~\ref{fig:Particle-number}(b), in the tFF state, the number density distributions of the two hyperfine states in the closed channel ($|g\uparrow\rangle$ and $|e\downarrow\rangle$) are apparently different and asymmetric with respect to zero momentum, which distinguishes the tFF state from all the other states. In fact, as the pairing order parameters are relatively small in the tFF state, the density distributions of atoms are close to those of a noninteracting gas, which, under the current parameters, features a Fermi surface lying below the states of the open channel $(|g\downarrow\rangle$, $|e\uparrow\rangle$). Thus, in the tFF state, the open channel is nearly unoccupied at zero temperature, while the occupations of the states in the closed channel $(|g\uparrow\rangle$, $|e\downarrow\rangle$) are different due to the Zeeman shift $\delta_c$. As a comparison, for the trivial FF state, the pairing order parameters are large compared with $\delta$, which is the typical energy offset between states in the open and the closed channel. As a consequence, the difference between the density distributions is negligibly small. Such a qualitative difference in number density distributions can serve as an indirect signal for the tFF state.

\section{Summary}
\label{sec:summary}
We have shown that by engineering the atom-laser couplings in alkaline-earth-metal-like atoms, a topological FF state can be stabilized in an ultracold atomic gas near an orbital Feshbach resonance. As the nontrivial topology of the state is entirely encoded in the closed channel, the system can be regarded as the coexistence of a tFF and a trivial FF state in two separate but coupled channels. It is interesting to study the stability and properties of the system beyond the mean-field description, i.e., how the quantum fluctuations would affect the topology of the underlying system. Note that effects of SOC under the same cross-coupling scheme have recently been considered in Ref.~\cite{iemini_majorana_2017}, which focuses on the charge and spin correlations in a strictly one-dimensional system and is in contrast to the pairing correlation of a quasi-one-dimensional system considered in this work.

\section*{Acknowledgement}
We would like to thank Dongyang Yu, Fan Wu, Fang Qin, Shunyao Zhang for helpful discussions. This work is supported by the National Natural Science Foundation of China (Grant Nos. 11274009, 11374177, 11374283, 11421092, 11434011, 11522436, 11522545, 11534014, 11626436, 60921091), the National Key R\&D Program (Grant Nos. 2016YFA0300603, 2016YFA0301700), the NKBRP (2013CB922000), and the Research Funds of Renmin University of China (10XNL016, 16XNLQ03). W. Y. acknowledges support from the ``Strategic Priority Research Program(B)'' of the Chinese Academy of Sciences, Grant No. XDB01030200.

\bibliographystyle{apsrev4-1}
\bibliography{tex}

\end{document}